\documentclass[epj]{svjour}
\usepackage{graphicx}
\usepackage{amsmath}
\usepackage{amssymb}
\usepackage{paralist}
\usepackage{color}
\newcommand{\ve}{\ensuremath{\varepsilon}}
\newcommand{\vt}{\ensuremath{\vartheta}}
\begin{document}
%
\title{Excitation Waves on a Minimal Small-World Model}
\author{
  Thomas Isele\inst{1} \and 
  Benedikt Hartung\inst{1} \and 
  Philipp H\"{o}vel\inst{1,2,}\thanks{Corresponding author: Philipp H\"ovel (\email{phoevel@physik.tu-berlin.de})} \and 
  Eckehard Sch\"{o}ll\inst{1}
}
%
%
\institute{
  Institut f\"{u}r Theoretische Physik, Technische Universit\"{a}t  Berlin, Hardenbergstra{\ss}e 36, 10623 Berlin, Germany 
  \and 
  Bernstein Center for Computational Neuroscience, Humboldt-Universit{\"a}t zu Berlin, Philippstra{\ss}e 13, 10115 Berlin, Germany
}

\date{Received: date / Revised version: date}
%
\abstract{
We examine traveling-wave solutions on a regular ring network with one additional long-range link that spans a distance $d$.
The nodes obey the FitzHugh-Nagumo kinetics in the excitable regime. 
The additional shortcut induces a plethora of spatio-temporal behavior that is not present without it.
We describe the underlying mechanisms for different types of patterns: propagation failure, period decreasing, bistability, shortcut blocking and period multiplication.
For this purpose, we investigate the dependence on $d$, the network size, the coupling range in the original ring and the global coupling strength and present a phase diagram summarizing the different scenarios. Furthermore, we discuss the scaling behavior of the critical distance by analytical means and address the connection to spatially continuous excitable media.
\PACS{
      {64.60.aq}{Networks}   \and
      {89.75.Fb}{Structures and organisation in complex systems} \and
      {82.40.Ck}{Pattern formation in reactions with diffusion, flow and heat transfer} \and
      {47.54.-r}{Pattern selection; pattern formation} 
     } 
\keywords{
  Network dynamics \and
  {Traveling waves} \and
  {Spatio-temporal patterns} \and
  {Excitable systems}
}
} 
\maketitle
\section{Introduction}\label{sec:intro}


In recent years, dynamical systems coupled in complex network architectures have become an important paradigm for studying phenomena in many areas of applications \cite{ALB00,ALB02a,NEW03,BOC06a,KIT10,DAH12,LIZ12}.
Patterns originally related to reaction-diffusion systems have been studied on complex networks as well.
Examples are Turing patterns in general network topologies \cite{NAK10} as well as traveling-fronts and pulses on regular trees \cite{KOU12,KOU14}.

Regular lattices are simple networks that exhibit a periodic structure. 
Excitable systems on regular lattices coupled by the corresponding Laplacian can be understood as discrete approximations of excitable media.
The study of waves on discrete systems is of interest in itself, as propagation properties on these can differ from those on continuous media \cite{CAR00,CAR05a,CAR03,LAM00a}.

The influence of non-local coupling on wave propagation in excitable media has received much attention recently \cite{BAC14,SIE14,SIE14a,LOE14,SCH09c}.
A possible non-local modification of regular lattices is a so-called small-world topology \cite{MIL67,WAT98}, where long-range links change the architecture of the network by introducing shortcuts through the network.
Wave propagation on these has also been investigated \cite{ISE14,SIN07b,ROX04}.
Under certain circumstances, it is possible to summarize the effect these small-world topologies have in a mean-field approximation of a modified continuous reaction-dif\-fu\-sion system \cite{ISE14}.

In this work we are interested in modifications of the topology and their effects that are genuine to the network structure and cannot be transferred to a continuous system.
This is in contrast to studies of dynamics on small-world topologies where there is inherent randomness in the construction of the topology and thus, most results are given for an ensemble of small-world networks, e.g. with the same long-range link density \cite{ISE14,SIN07b,ROX04}.
It is the purpose of this paper to separate the specific geometric effects of the long-range links from the randomness introduced by the small-world topology and study systematically the spatio-temporal patterns that arise through the modification.
A minimal version of the Newman-Watts small-world model \cite{NEW03} serves these purposes.
It consists of a one-dimensional regular lattice with one additional long-range link.
If the distance $d$, which this link spans, is systematically varied, no randomness is involved in constructing the topology.
This long-range link is an exclusive feature of the combination of regular structure, which provides a notion of distance and which supports traveling waves, and the network nature of the system that allows to simply link two arbitrary nodes.

We will show that this small modification of the network topology facilitates an abundance of spatio-temporal patterns which are not present in the system without it.
Furthermore, we demonstrate that -- in order to control the selection of patterns -- changing the distance $d$ of the long-range link or changing the overall coupling strength $\kappa$ is sufficient.

The rest of this paper is structured as follows:
In Sec.~\ref{sec:model}, we introduce our network model and the dynamics under consideration.
In Sec.~\ref{sec:spat_temp_patt}, we present the different spatio-temporal patterns for varying $d$ and $\kappa$, and illustrate the mechanisms of their generation.
In Sec.~\ref{sec:phase_diags}, we give a survey plot of all spatio-temporal behavior for fixed network size $N$ and coupling range $k$.
There, we also investigate how the pattern selection with respect to $d$ and $\kappa$ depends on $N$ and $k$ and extract the scaling behavior for three different dynamical regimes.
In Sec.~\ref{sec:analytics} we present some analytic calculation which deepens the understanding of the mechanisms explained before.
Section~\ref{sec:conclusion} finally concludes this manuscript.

\section{Model}\label{sec:model}
In the present study we will examine excitable activator-inhibitor dynamics on a network that consists of a regular ring with $N$ nodes.
In this regular ring, each node is coupled to its $k$ neighbors to the left and to the right.
Furthermore, one additional link is inserted between one pair of nodes that are a distance $d\in\{k+1,k+2,...,N/2\}$ apart.
We are considering only unweighted, undirected links.
The adjacency matrix thus reads as follows
\begin{align}
  A_{ij}   &=  \sum_{\mu=1}^k \left(\delta_{i,j+\mu} + \delta_{i,j-\mu}\right) + \delta_{i,0}\delta_{j,d} + \delta_{i,d}\delta_{j,0}\ , \label{eq:adjacency_matrix}
\end{align}
where all indices are to be understood modulo $N$, $\delta_{i,j}$ is the Kronecker symbol and the additional link connects nodes 0 and $d$.

\begin{figure}
\resizebox{0.4\columnwidth}{!}{%
  \includegraphics{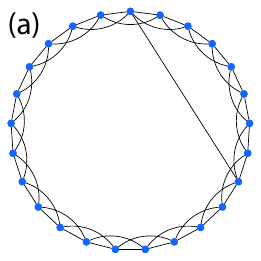}
}%
\resizebox{0.6\columnwidth}{!}{%
  \includegraphics{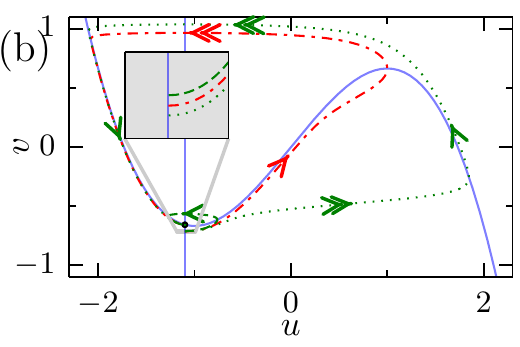}
}
\caption{(a) Scheme of a network with $N=25$, $k=2$, $d=8$.
         (b) Phase portrait of the FitzHugh-Nagumo system, including nullclines (solid, blue), canard trajectory
(dash-dotted, red), a sub-threshold excitation (dashed, green) and a super-threshold excitation (dotted, green).
         The inset shows a blow-up of the region near the fixed point.
         Parameters: $\ve=0.04,\,\beta=-1.1\,$.}
\label{fig:system_illustration}       
\end{figure}

A scheme of such a network is given in Fig.~\ref{fig:system_illustration}(a) for $N=25,\,k=2$ and $d=8$.
Note that this network is invariant with respect to reversing the order of the nodes and subsequently shifting the node indices by $d$.
Also, a network with an additional link between nodes 0 and  $d>N/2$ can be tranformed to a network with an additional link from node 0 to node $\tilde d = N-d<N/2$ using only the shift symmetry operation.
This is the reason for restricting $d\in\{k+1,k+2,...,N/2\}$.
The shift symmetry is also used in all space-time plots to place the additional link between nodes that do not lie at the edges of the plot. See, for instance, Fig.~\ref{fig:unperturbed_timeseries}(b), which will be discussed in detail later. There, the shortcut is added between nodes 20 and 70 instead of nodes 0 and 50.

Note that any unidirectionally traveling wave solution breaks the symmetry $i\to d-i\,\text{mod}\,N$ and thus allows us to distinguish the `first' and `second' end of the additional link.
Additionally, in Sec.~\ref{sec:bistability}  we will present a mechanism breaking the symmetry $d\to\tilde d=N-d$ such that spatio-temporal patterns for $d$ and $\tilde d$ are different.
We will also present a second mechanism in Sec.~\ref{sec:shortcut_block} that reinstalls the symmetry between spatio-temporal patterns for $d$ and $\tilde d$.

The considered network model is a minimal version of the Newman-Watts small-world model \cite{NEW03,WAT98}, in which a certain number of additional random links is inserted into such a regular ring network. Wave propagation in Newman-Watts small-world networks has been studied in Ref.~\cite{ISE14} with focus on the collective effects of a large number of such links, where that configuration can be approximated by a mean-field. 
The reason for choosing the network with just one additionally inserted link is to study in more detail and systematically the effect of shortcuts through the ring.

The nodes of the network follow the FitzHugh-Nagumo dynamics \cite{FIT61,NAG62} and are diffusively coupled via the
activator.
Thus, the full equations for the dynamics read as follows:
\begin{subequations}
\label{eq:dynamics}
\begin{align}
  \dot u_i    &= u_i - \frac{u_i^3}{3} - v_i  + \kappa \sum_{j=1}^N A_{ij}\left(u_j-u_i\right) \label{eq:dynamics_u} \\
  \dot v_i    &= \ve( u_i - \beta ), \quad i=0,...,N-1                               \label{eq:dynamics_v}
\end{align}
\end{subequations}
where $\ve$ and $\beta$ are the time-scale separation and threshold parameter, respectively,
$\kappa$ is the coupling strength, $u$ is the (fast) activator and $v$ is the (slow) inhibitor variable.
Throughout this work, we will fix the values for these parameters at $\beta=-1.1$ and $\ve=0.04$.
In this regime, the dynamics has one stable fixed point. 
The so-called canard trajectory serves as a threshold trajectory that separates initial conditions which perform a small
excursion in phase space from those that induce a large excursion along both stable branches of the $u$-nullcline. 
A phase plane plot is shown in Fig.~\ref{fig:system_illustration}(b) for illustration. The red (dash-dotted), green
dashed and green dotted curves depict trajectories of three difference initial conditions leading to the canard
trajectory, a sub-threshold excitation and a super-threshold excitation, respectively.

Note also that due to the coupling being only in the activator, the stable fixed point of the local dynamics becomes a
\emph{stable} homogeneous fixed point of the entire system.
This follows from a simple calculation of the eigenvalues of the system Eqs.~\eqref{eq:dynamics}, for details see
\cite{ISE14}.

\section{Results}\label{sec:results}
\subsection{Spatio-Temporal Patterns}\label{sec:spat_temp_patt}
In this section we systematically discuss and analyze the emerging patterns by varying $d$ and $\kappa$.
On the regular ring without the additional link, successive mutual excitation of the nodes leads to the emergence of a
stable traveling wave.
By choosing the initial conditions carefully, a prescribed direction of propagation can be chosen.
We choose the initial conditions such that a single excitation pulse traveling in direction of ascending node indices is
generated.

Depending on the global coupling strength $\kappa$ in Eqs.~\eqref{eq:dynamics}, the propagation speed of the pulse, as
well as the pulse width differ.
For each combination of $N$ and $k$, there is a maximum coupling strength, given by
$\kappa_\text{max}=\frac{N^2}{q(k)L_\text{min}(\ve,\beta)^2}$ with $q(k)=\frac 1 6 k(k+1)(2k+1)$ and the minimum domain
size $L_\text{min}$ allowing for stable wave propagation in the spatially continuous reaction-diffusion system
$\partial_{t} u = u-\frac{u^3}{3}-v + \Delta u,\ \partial_{t} v=\ve(u-\beta)$, where $\Delta$ denotes the diffusion
operator (second derivative with respect to space, $\Delta=\partial_{xx}$) subject to  periodic boundary conditions
$u(t,0)=u(t,L)$, $v(t,0)=v(t,L)$. 
For our choice of parameters, we find $L_\text{min}(\ve=0.04,\beta=-1.1)\approx30.756$. 
At this maximum coupling strength, the branch of stable traveling waves is destabilized by a torus bifurcation.
This has been discussed in \cite{ISE14}.

Due to the discrete nature of the system, there is also a minimum value of the coupling strength $\kappa_\text{min}$
that only depends on $k,\ \ve$ and $\beta$ but not on the system size $N$.
At this value, the branch of stable traveling waves meets the branch of unstable traveling waves in a saddle-node
bifurcation. 
A more detailed account on these matters including values of $\kappa_\text{min}(k)$ is provided in Ref.~\cite{ISE14}
(cf. Table~\ref{tab:kappa_vals}).

\begin{table}
\centering
\begin{tabular}{r||l|l|l}
  $k$ &  $\kappa_\text{min}$ & $\kappa_1$         & $\kappa_2$  \\
\hline
1     &  $0.0324$    & $0.0339$    & $0.0359$   \\
2     &  $0.0233$    & $0.0235$    & $0.0481$  \\
3     &  $0.0169$    & $0.0170$    & $0.5890$   \\
\end{tabular}
\caption{Approximate transition values $\kappa_\text{min},\,\kappa_1$ and $\kappa_2$ for different nearest neighbor
numbers $k$. Parameters: $\ve=0.04,\,\beta=-1.1$. See also \cite{ISE14}.\label{tab:kappa_vals}}
\end{table}

Based on network Eq.~\eqref{eq:adjacency_matrix}, we will examine the effects of one additional link on the traveling
wave solution present on the regular ring.
To this end, we choose as initial conditions the snapshot of a stable traveling pulse from the regular ring with the
peak before the first end of the additional link. Then, we numerically integrate the initial value problem on the system
including that additional shortcut link.
We use a Runge-Kutta Fehlberg 4(5) scheme with adaptive stepsize.

A traveling wave solution on the network without the additional link is a periodic solution that has a well-defined
period $T_0(\kappa)$.
In the following, we discuss changes of this periodic solution for networks including the additional link.
In short, we observe either (a) an ongoing spatio-temporal activity or (b) relaxation into the stable homogeneous steady
state (propagation failure).
In almost all cases where ongoing spatio-temporal activity emerges, we are able to measure the period of the emerging
solution $T_d$, where $d$ is the distance which the additional link spans. 
To quantify the spatio-temporal behavior of these solutions, we measure the relative period defined by $\vt_d (\kappa):=
T_d(\kappa) / T_0(\kappa)$. 

For this purpose, we will present examples for typical spatio-temporal patterns. In all examples, we consider $N=150$
and $k=2$, which yields $\kappa_{\text{max}}\approx 4.75$ and $\kappa_{\text{min}} \approx 0.0233$.
Next we will discuss the observed scenarios in greater detail one by one and then summarize our findings in a phase
diagram of $\vt_d$
vs. $\kappa)$.

\subsubsection{Unperturbed wave, additional link has no influence}\label{sec:unpert_wave}
\begin{figure*}
\resizebox{\textwidth}{!}{%
  \includegraphics{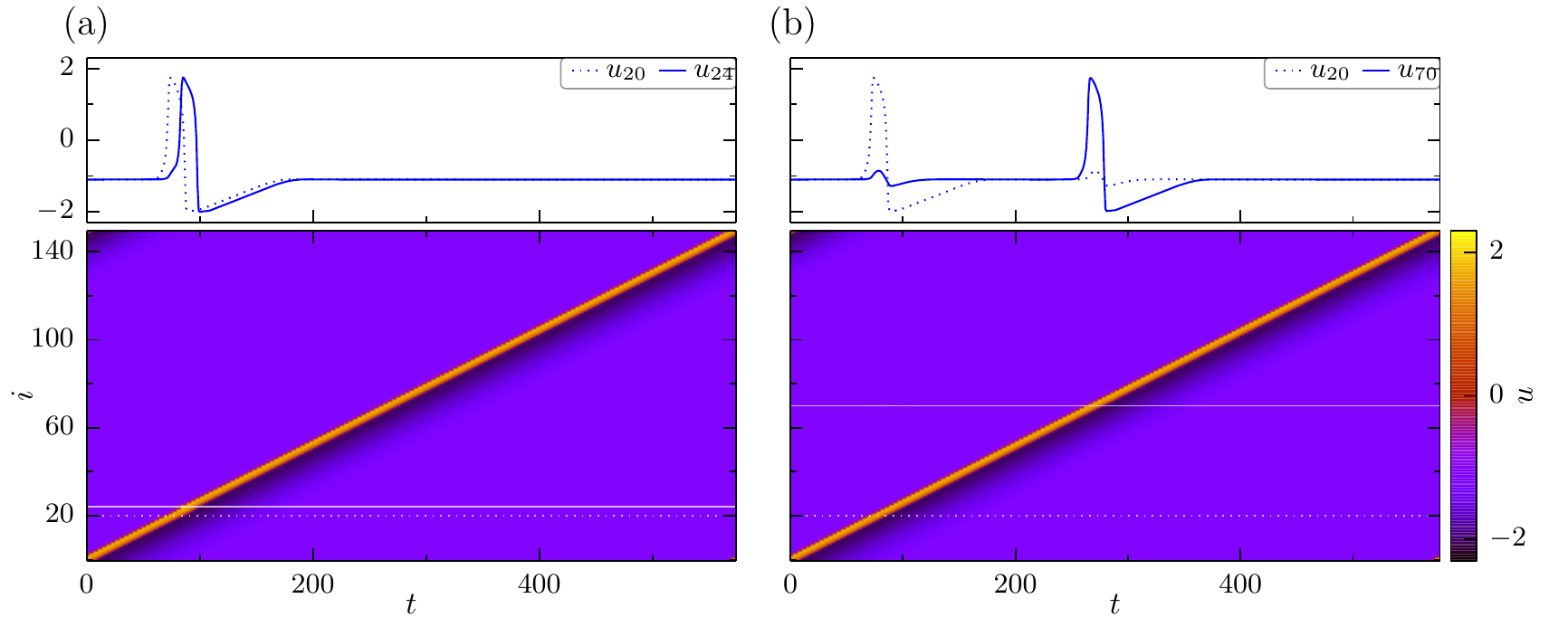}%
}%
\caption{Color-coded activator level $u_i(t)$ as space-time plot (lower panel) and timeseries (upper panel) (a)
$u_{20}(t)$, $u_{24}(t)$ and (b) $u_{20}(t)$, $u_{70}(t)$ over one period of a traveling wave solution of
Eqs.~\eqref{eq:dynamics} for
(a) $\kappa=0.03,\ d=4$,  resulting period $T_4=574.0\   (\vt_4(0.03)=0.996)$ and
(b) $\kappa=0.03,\ d=50$, resulting period $T_{50}=577.2\ (\vt=1.001)$ (cf. Sec.~\ref{sec:unpert_wave}).
Parameters:  $N=150,\ k=2,\ \beta=-1.1,\ \ve=0.04$. The additional link is placed in panel (a) between nodes 20 and 24
and in panel (b) between nodes 20 and 70, marked by horizontal lines in the space-time plot with linestyles
corresponding to the respective timeseries.
}
\label{fig:unperturbed_timeseries}       
\end{figure*}

If the additional link covers a distance that is only slightly larger than the coupling range of the regular ring, i.e.,
$d\gtrsim k$, it has no significant influence on the traveling wave.
An example for this case is shown in Fig.~\ref{fig:unperturbed_timeseries}(a) for a coupling strength $\kappa=0.03$.

As another example, we show the solution for the network with an additional link with $d=50$ between nodes 20 and 70 in
Fig.~\ref{fig:unperturbed_timeseries} and $\kappa=0.03$.
In the timeseries of $u_{20}$ and $u_{70}$, we see that the wave triggers a small sub-threshold excitation at the
respective other node. 
The periods are $T_0(0.03)=576.2$ and $T_{50}(0.03)=577.2$ so that $\vt_{50}(0.03)=1.001$ which indicates that the
additional link slows down the wave a little.
The influence is very small for this particular configuration.

\subsubsection{Propagation failure I,  low $\kappa$: Direct failure}\label{sec:PD_direct}
\begin{figure*}
\resizebox{1.0\textwidth}{!}{%
  \includegraphics{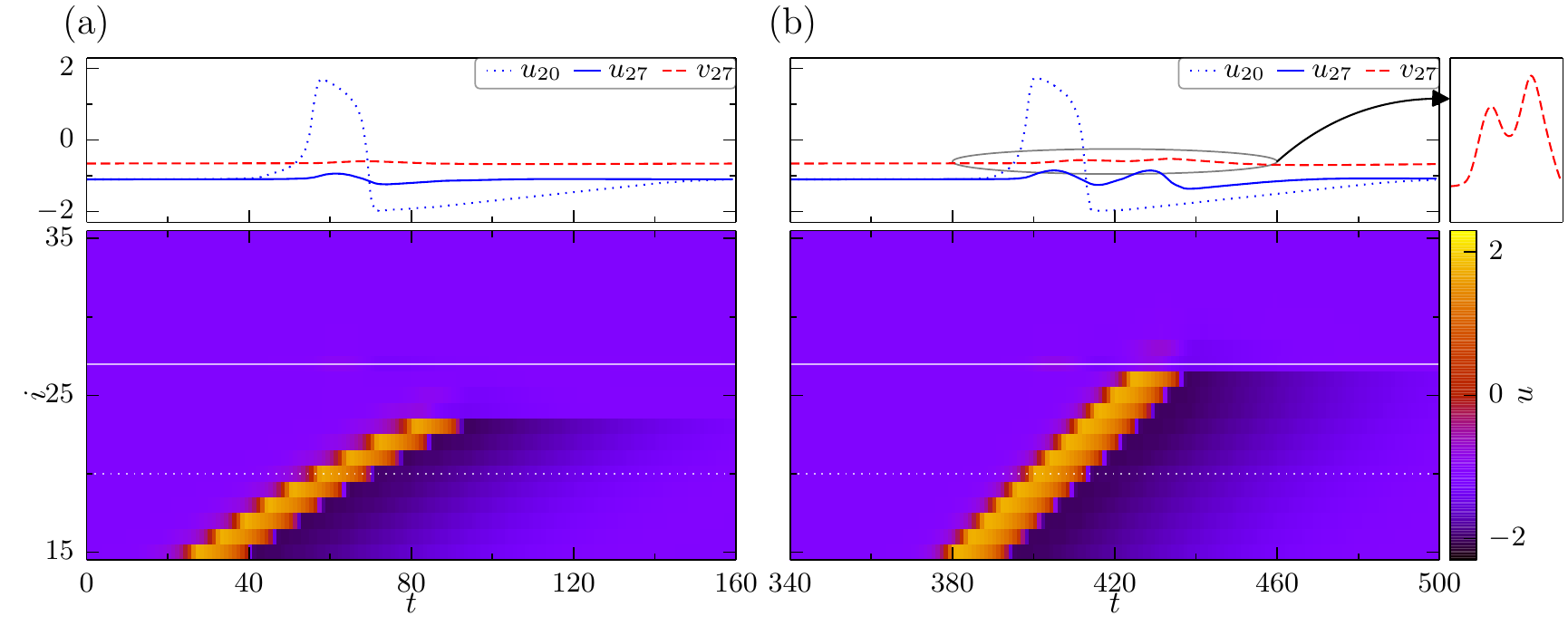}
}
\caption{Color-coded activator level $u_i(t)$ as space-time plot (lower panel) and timeseries (upper panel) of
$u_{20}(t),\ u_{27}(t)$, and $v_{27}(t)$ for 
         (a) $\kappa=0.02335,\ d=7$, resulting in direct failure (cf. Sec.~\ref{sec:PD_direct}) and 
         (b) $\kappa=0.03,\ d=7$, resulting in inhibitor-mediated failure (cf. Sec.~\ref{sec:PD_inh_med}).
         In the very upper right inset, a blow-up of $v_{27}$ for the encircled region is shown.
         Other parameters as in Fig.~\ref{fig:unperturbed_timeseries}.
         An additional link is placed between nodes 20 and 27, which are marked by horizontal lines in the space-time
plot with linestyles corresponding to the respective timeseries.
}
\label{fig:direct_and_indirect_failure}       
\end{figure*}

When the coupling strength $\kappa$ is below a certain value $\kappa_1(k)$, one additional link, spanning any distance
larger than $k$, is enough to suppress propagation.
Values of $\kappa_1(k)$ are determined by numerical simulation for $k=1,2,3$ and are listed in
Table~\ref{tab:kappa_vals}.
For such low coupling strengths, the propagation takes place in a `saltatory' way, i.e., the transitions to excitation
of subsequent nodes are well separated in time. (See Fig.~\ref{fig:direct_and_indirect_failure}(a)).

In the case of direct failure, the additional coupling of one node to any other node (which will be in -- or very close
to -- the fixed point) is enough to prevent its full excitation.
Instead it performs a sub-threshold excitation in which the activator does not rise high enough to excite the next node
on the ring.
This mechanism is depicted in Fig.~\ref{fig:direct_and_indirect_failure}(a) and has also been described in
Ref.~\cite{ISE14}.

However, the failure can occur a couple of nodes after the first node in which the additional link ends.
In that case, the times between the excitation peaks of the subsequent node becomes larger.
This has been described (in the context of coupling strengths below $\kappa_{\text{min}}$) in Ref.~\cite{BOO95}.

\subsubsection{Propagation failure II, low $\kappa$: Inhibitor mediated failure}\label{sec:PD_inh_med}
When the coupling strength $\kappa$ exceeds $\kappa_1(k)$, propagation failure can still occur but not at any distance
that the additional link spans.

In passing the node at one end of the additional link, the high activator level will cause an excitation at the other
end of it.
Below a certain coupling strength $\kappa_2(k)$ (determined numerically and listed in Table~\ref{tab:kappa_vals}), this
excitation stays sub-threshold and does not lead to an excitation of a secondary wave.
Nevertheless, the sub-threshold excitation leads to a rise in inhibitor level at the node at the other end of the
additional link.
If the distance the link spans is small enough such that the wave will arrive at the other end, when the inhibitor is
still excited, the wave will terminate.
If the distance is larger, the inhibitor level will have relaxed enough to let the wave pass, in this case the
additional link exerts no significant influence (cf. Sec.~\ref{sec:unpert_wave}).
This mechanism is shown in Fig.~\ref{fig:direct_and_indirect_failure}(b) and has also been described in
Ref.~\cite{ISE14}.

\subsubsection{Period decreasing}\label{sec:period_decr}
Above a value $\kappa_2(k)$ of the coupling strength, a super-threshold excitation can be triggered via the additional
link.
In principle, such an excitation always leads to the generation of a pair of excitation waves traveling in opposite
directions at the second end of the link.
One wave of this pair travels in the opposite direction towards the original wave and at some point both collide and
annihilate each other.
The other part of this pair continues traveling along the network until it again reaches the node at the first end of
the link and the pattern repeats.

This pattern leads to a decrease of the effective size of the regular ring by approximately $d$, as the pattern repeats
when one half of the secondary wave pair generated at the second end of the additional link has reached the first end of
the link. 
Thus, the wave effectively skips the part of the network where the two counterpropagating pulses annihilate.
As a consequence, the larger $d$ becomes, the smaller becomes $\vt_d(\kappa)$.
Approximately we have $\vt_d(\kappa)\approx 1-  d/N$.
becomes red.

\subsubsection{Bistability}\label{sec:bistability}
Linked to the phenomenon of period-decreasing is the breaking of the symmetry of Eq.~\eqref{eq:dynamics} that maps an
additional link with distance $d$ onto one with distance $\tilde d=N-d$ (and shifting the node indices).

If different initial conditions from the ones mentioned above are chosen, namely when the peak of the wave is placed
before the second end of the additional link, the part where the counterpropagating pulses annihilate comprises $N-d$
nodes instead of $d$ thus making the part that the wave effectively skips larger and thus $\vt_d(\kappa)\approx
1-\frac{N-d}N=\frac d N < 1-\frac d N$.

\begin{figure*}
\resizebox{1.0\textwidth}{!}{%
  \includegraphics{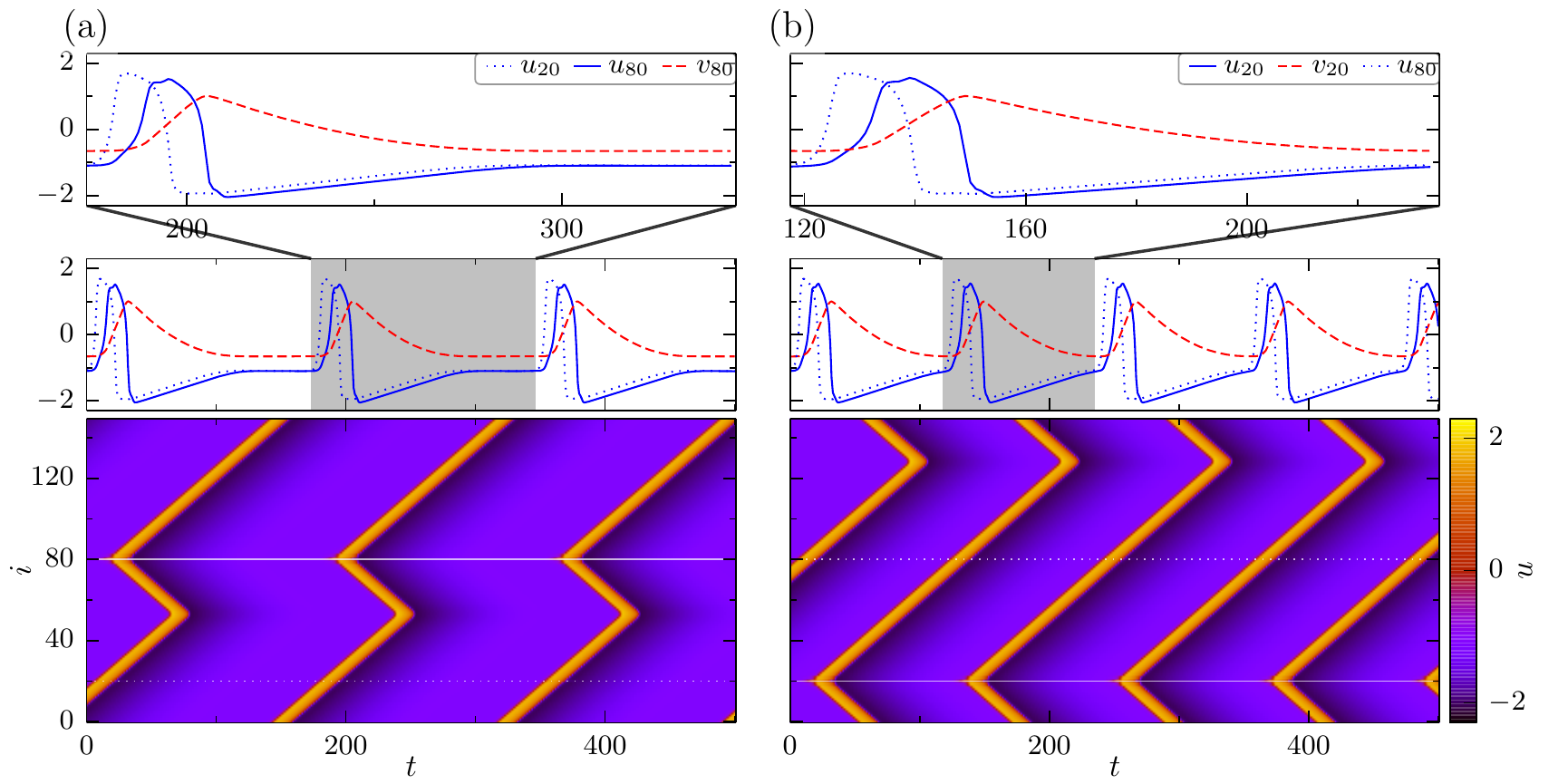}
}%
\caption{Color-coded activator level $u_i(t)$ as space-time plot (bottom panel) and timeseries of activator and
inhibitor levels at both ends of the additional link, marked by horizontal lines in the space-time plot.
         (a) $\kappa=0.08,\ d=60$, initial wave peak placed in the large part of the ring, leading to annihilation in
the small part of the ring, resulting period $T_{60}(0.08)=173.24\ (\vt_{60}(0.08)=0.64)$ and
         (b) $\kappa=0.08,\ d=60$, initial wave peak placed in the small part of the ring, leading to annihilation in
the large part of the ring, resulting period $T_{60}(0.08)=117.41\ (\vt_{60}(0.08)=0.43)$ (cf.
Sec.~\ref{sec:bistability}).
         Top panels show one period, middle and bottom panels show same absolute timespan for both (a) and (b).
         Other parameters as in Fig.~\ref{fig:unperturbed_timeseries}.
         }
\label{fig:period_decrease_and_bistability}       
\end{figure*}

An example for this behavior is shown in Fig.~\ref{fig:period_decrease_and_bistability}. 
In that example, we choose $N=150,\ k=2,\  d=60,\ \kappa=0.08$ and therefore, by the approximation mentioned above, we
expect $\vt_60(.008)\approx 1-\frac{60}{150}=0.6$ and $\vt_90(0.08)\approx\frac{60}{150}=0.4$. The numerically
determined values are both a little larger (0.64 and 0.43), thus accounting for the time it takes to trigger a wave
through the single additional link instead of locally through two links ($k=2$).

\subsubsection{Shortcut blocking }\label{sec:shortcut_block}
After undergoing the excitation, a node is left with increased inhibitor level, which slowly relaxes to the fixed point
on the `left' branch of the $u$-nullcline (see Fig.~\ref{fig:system_illustration}(b)).
All nodes that are in this state form the refractory tail of the wave.

The triggering of a super-threshold excitation through the additional link can be suppressed by the refractory tail when
it extends back to the other end of the additional link.
Then there is an increased inhibitor level which prohibits a super-threshold excitation. 
For an analytic calculation to determine the critical value of the inhibitor see Sec.~\ref{sec:analytics}. 

When shortcut blocking occurs for the link distance $d$ but not for the link distance $N-d$, the bistability between the
two distances is lifted and thus the symmetry recovered.
The reason is that initial conditions with the wave peak before either end of the additional link will lead to a
solution with decreasing period (see \ref{sec:period_decr}) in which the counterpropagating pair annihilates in the
smaller part of the ring. 
Evaluating all link distances (up to $N-k$), one can see this phenomenon at a fixed value of $\kappa$ as a sudden
transition of $\vt_d(\kappa)$ in $d$.
Note also that the refractory tail becomes larger as $\kappa$ increases and thus the transition happens at lower values
of $d$ (because the refractory tail has to bridge $N-d$ nodes).
Therefore, the point of transition can be regarded as a measure of the size of the wave. 

\begin{figure*}
\resizebox{1.0\textwidth}{!}{%
  \includegraphics{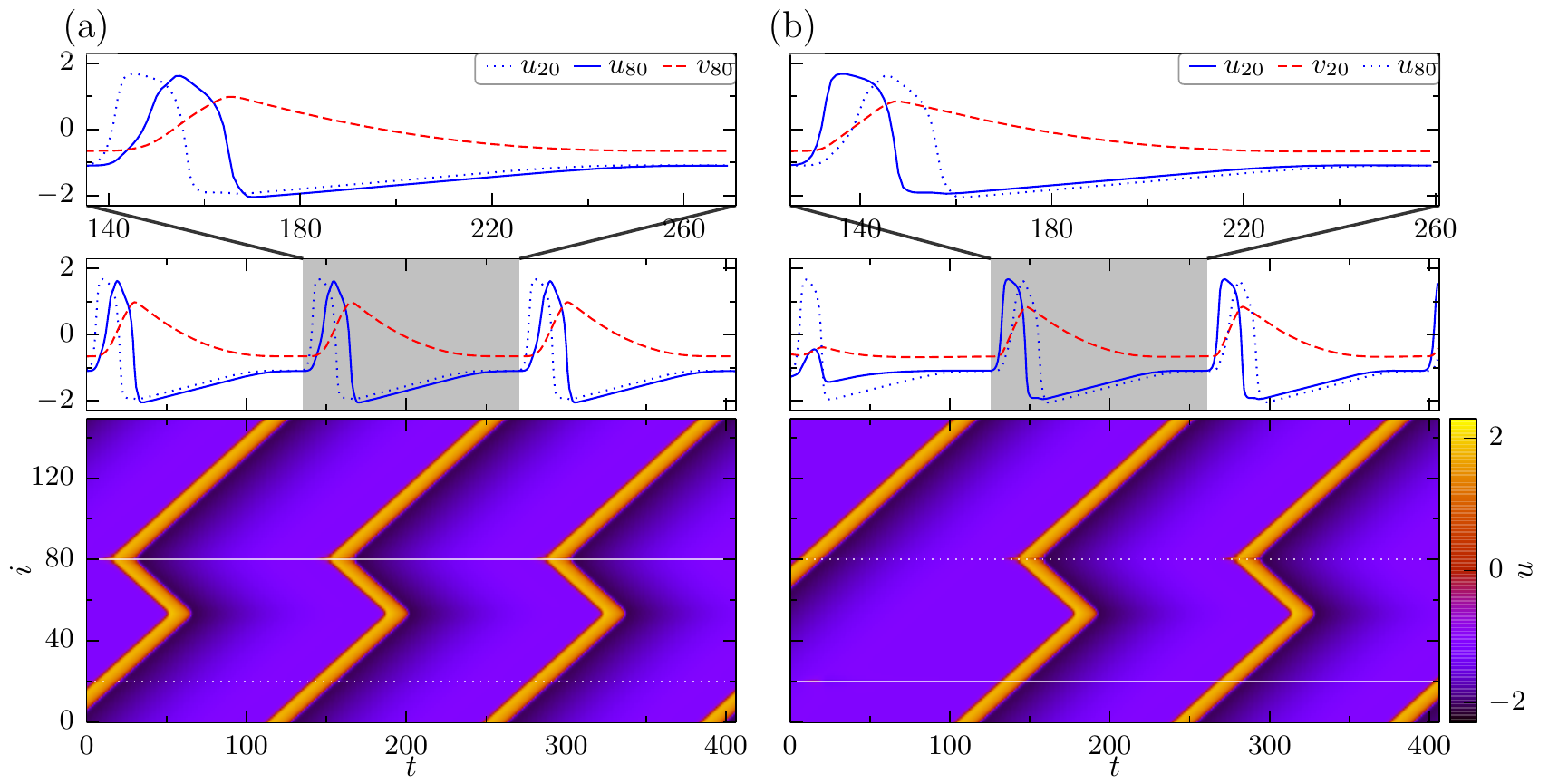}
}%
\caption{Color-coded activator level $u_i(t)$ as space-time plot (bottom panel) and timeseries of activator and
inhibitor levels at both ends of the additional link, marked by horizontal lines in the space-time plot.
         (a) $\kappa=0.12,\ d=60$, initial wave peak placed in the large part of the ring and
         (b) $\kappa=0.12,\ d=60$, initial wave peak placed in the small part of the ring (cf.
Sec.~\ref{sec:shortcut_block}).
         Differently from the situation in Fig.~\ref{fig:period_decrease_and_bistability}, the secondary wave pair is
always triggered at the same end of the additional link, after transients are gone. 
         Both solutions have the same period $T_{60}(0.12)=135.42\ (\vt_{60}(0.12)=0.65)$. 
         Top panels show one period of each solution, middle and bottom panels show same absolute timespan for both (a)
and (b).
         Other parameters as in Fig.~\ref{fig:unperturbed_timeseries}.}
\label{fig:shortcut_block}       
\end{figure*}

Exemplary timeseries for this behavior are shown in Fig.~\ref{fig:shortcut_block} at a coupling strength $\kappa=0.12$.
The setup is exactly the same as for Fig.~\ref{fig:period_decrease_and_bistability}, except that the overall coupling
strength is set to a higher value of $\kappa=0.12$.
In both Fig.~\ref{fig:shortcut_block}(a) and (b), the annihilation of the counterpropagating wave pair takes place in
the smaller part of the ring.

For values of $d$ above $N/2$, the shortcut will always be blocked.
For values of $\kappa$ higher than a certain threshold, period decreasing does not happen anymore.
However, other interesting phenomena start to occur.

\subsubsection{Period multiplication, complex spatio-temporal behavior}\label{sec:period_multiplication}
\begin{figure*}
\resizebox{1.0\textwidth}{!}{%
  \includegraphics{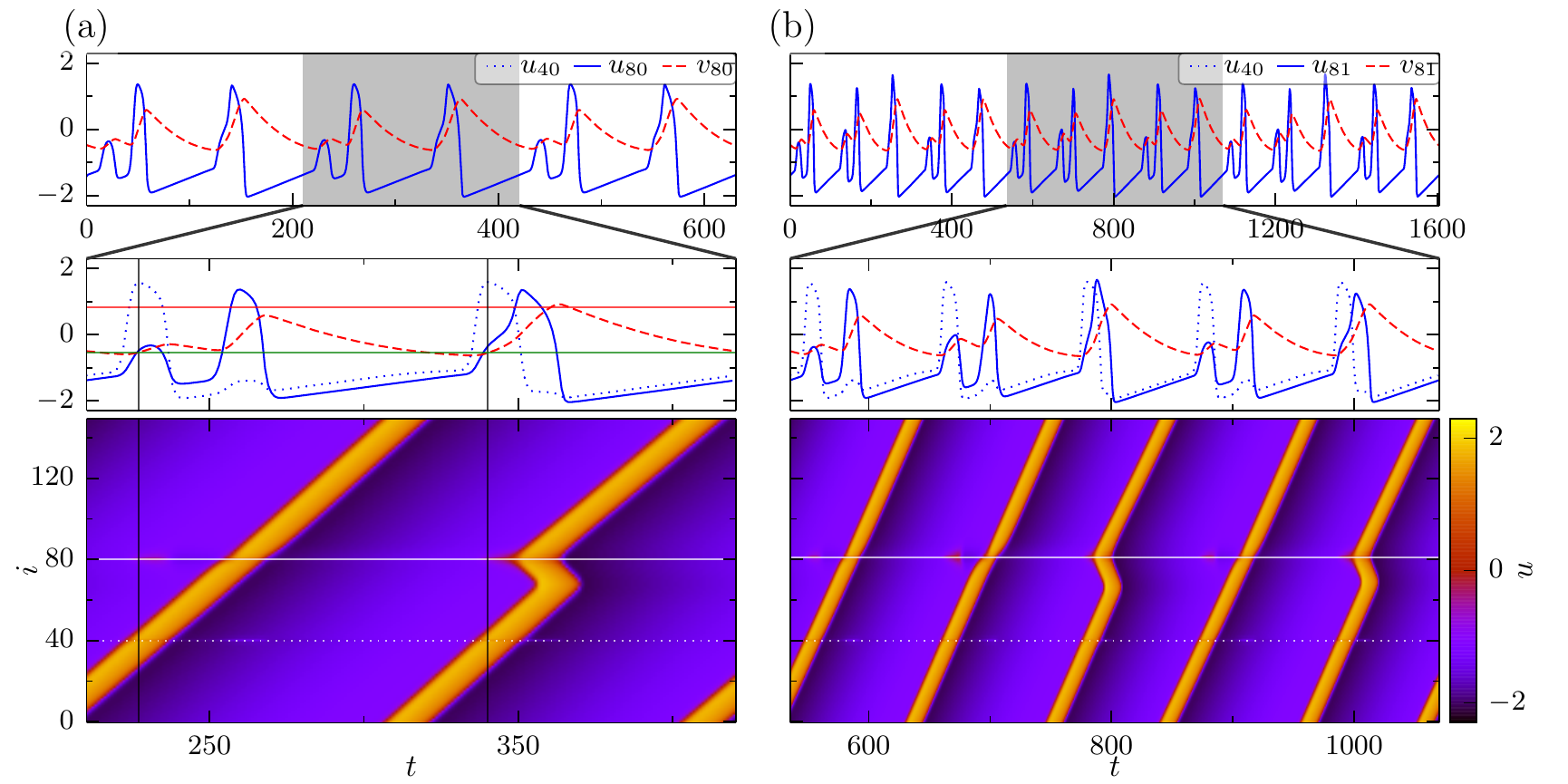}
}%
\caption{Color-coded activator level $u_i(t)$ as space-time plot (bottom panel) and timeseries of activator and
inhibitor levels at both ends of the additional link, marked by horizontal lines in the space-time plot.
         (a) $\kappa=0.37,\ d=40$, with resulting period-2 solution, $T_{40}(0.37)=210.2\ (\vt_{40}(0.37)=1.92)$ and 
         (b) $\kappa=0.37,\ d=41$, with resulting period-5 solution, $T_{41}(0.37)=535.2\ (\vt_{41}(0.37)=4.89)$.
         Top panels show three full periods, middle and bottom panels show one period.
         In the middle panel of (a) are marked by lines: Times of peak values of $u_{40}$ (gray), $v_{80}$ at the first
of these times (green), usual maximum inhibitor level when the wave passes through a node regularly (red).
         Detailed description in Sec.~\ref{sec:period_multiplication}.
         Other parameters as in Fig.~\ref{fig:unperturbed_timeseries}.
         }
\label{fig:period_multiplication}       
\end{figure*}
For some network configurations $(N,k)$ there can be complex behavior at certain values of $d$ and $\kappa$.
The relative period $\vt$ suddenly takes (very close to) integer values larger than 1, meaning that the spatio-temporal
pattern will repeat only after multiple passages of the wave through a fixed node.

We exemplarily choose a period-2 and a period-5 solution in Fig.~\ref{fig:period_multiplication} to illustrate this type
of behavior.
As we are still in the shortcut blocking regime and the coupling strength $\kappa$ is such that the wave size with
respect to shortcut blocking is larger than $N/2$, the first end of the additional link will always be blocked.

These solutions can be regarded as an interplay between shortcut blocking of the second link end and secondary wave pair
generation.
We find that the inhibitor level at the second end of the additional link --especially at the times when the activator
at the first end reaches the maximum value-- plays a key role in understanding the behavior.

In the period-2 example, the sequence of events might be described as follows:
\begin{inparaenum}[(a)]
\item Shortcut blocking, no secondary wave pair but a sub-threshold excitation is triggered at the second end of the
additional link.
\item The wave regularly arrives at the second end of the link, the peak is a little smaller compared to other nodes
because the inhibitor level is still a little raised from the sub-threshold excitation.
\item For that reason, the inhibitor does not rise as much as it would normally.
\item The wave arrives at the first end of the shortcut again.
Because the inhibitor level at the second end has not reached the usual maximum level when the wave has passed through,
shortcut blocking does not happen this time.
\item A secondary wave pair is triggered at the second end of the node. 
Now the inhibitor level at this end of the additional link rises even a little above the usual maximum level.
The reason is that the triggering of the secondary wave pair takes place through only one link as compared to two local
links ($k=2$), so $\dot v > 0$ for a longer time than usually. 
\item One half of the pair cancels out with the original wave, the other keeps on traveling until it reaches the first
end of the additional link again and the pattern repeats.
\end{inparaenum}

A similar but more complicated chain of events leads to the period-5 behavior also shown in
Fig.~\ref{fig:period_multiplication}.
Other periods (3,4,6,...) have also been observed in our simulations.

\subsubsection{Propagation failure III, intermediate $\kappa$: Complex mechanisms}\label{sec:PD_complex}

In the parameter ranges of $d$ and $\kappa$ where we observe period multiplication behavior, regimes of propagation
failure are always close.

The chain of events leading to propagation failure in this parameter regime, however, is more complex.
As we are still in the shortcut blocking regime, the first end of the additional link is always blocked and so the
failure always happens at the second end.
Like in the inhibitor-mediated failure for very small coupling strengths, the inhibitor level at the second end of the
additional link plays a key role here.
In this parameter regime, the wave will undergo several passages with chains of events similar as in
Sec.~\ref{sec:period_multiplication}.
In the end, however, the inhibitor concentration is too high to support passing-through of the wave and it stops at the
second link end.

\begin{figure*}
\resizebox{1.0\textwidth}{!}{%
  \includegraphics{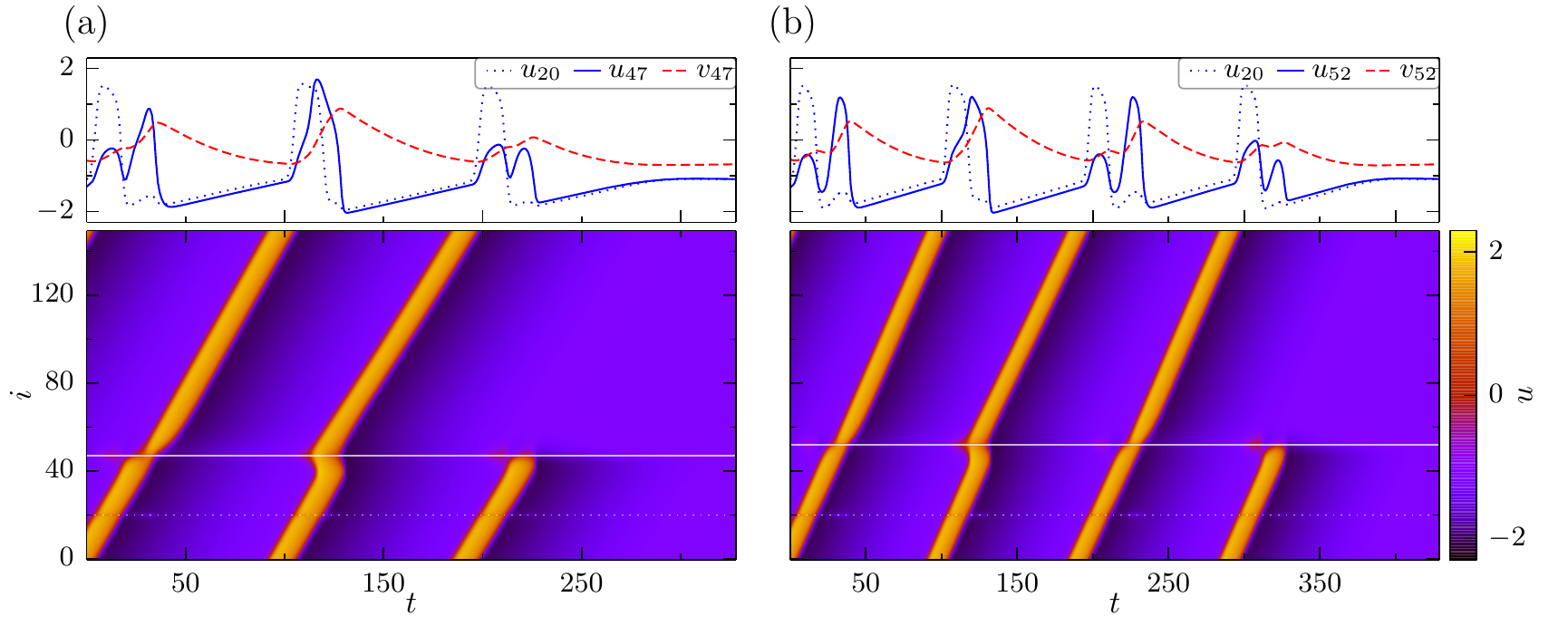}
}%
\caption{Color-coded activator level $u_i(t)$ as space-time plot and timeseries of activator and inhibitor levels at
both ends of the additional link, marked by horizontal lines in the space-time plot.
         (a) $\kappa=0.5,\ d=27$ and
         (b) $\kappa=0.5,\ d=32$.
         Both solutions are examples for complex mechanisms of propagation failure at an intermediate value of the
coupling strength $\kappa$ (cf. Sec.~\ref{sec:PD_complex} for more details).
         Other parameters as in Fig.~\ref{fig:unperturbed_timeseries}.}
\label{fig:PD_complex}       
\end{figure*}

In Fig.~\ref{fig:PD_complex}, we show two solutions where the wave stops only after several passages around the network.

\subsubsection{Propagation failure IV, high $\kappa$}\label{sec:PD_high_kappa}
\begin{figure*}
\resizebox{1.0\textwidth}{!}{%
  \includegraphics{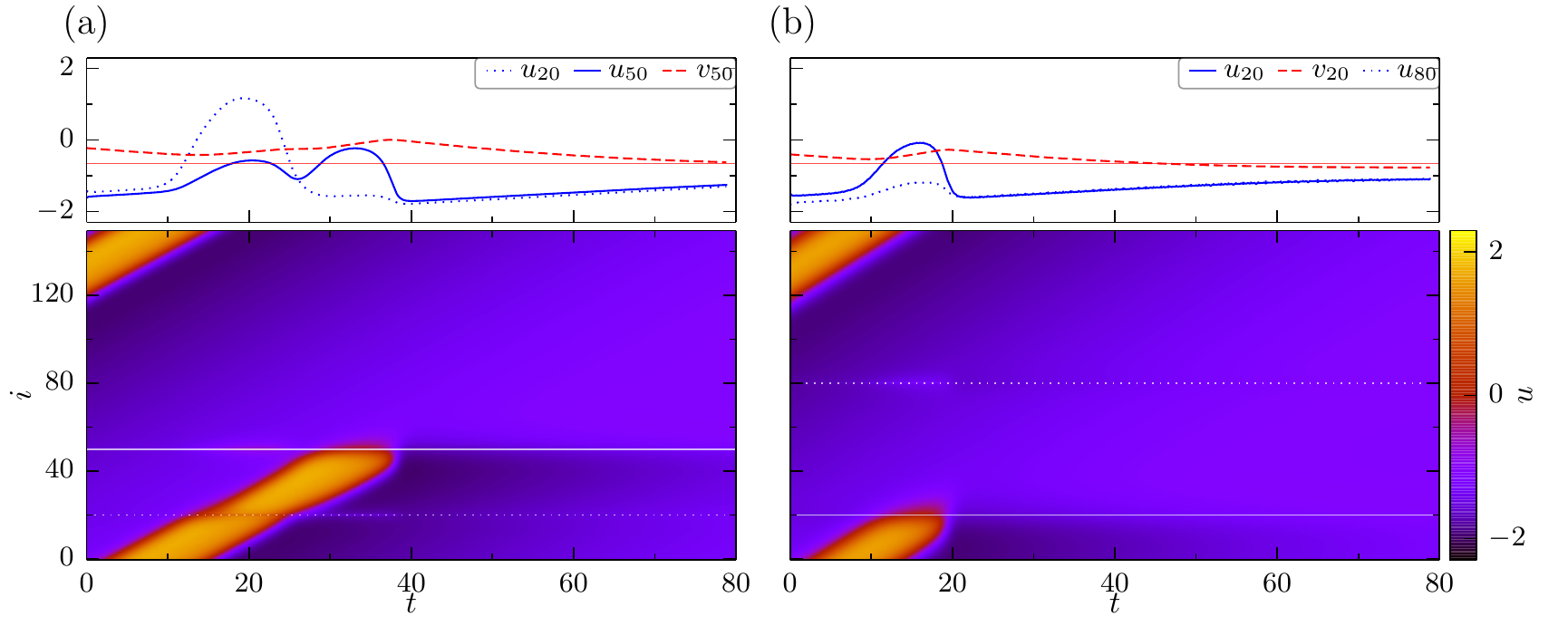}
}%
\caption{Color-coded activator level $u_i(t)$ as space-time plot (bottom panel) and timeseries of activator and
inhibitor levels at both ends of the additional link, marked by horizontal lines in the space-time plot.
         (a) $\kappa=1.5,\ d=30$ and
         (b) $\kappa=3.0,\ d=60$.
         Both solutions are examples for quenched propagation at high values of the coupling strength $\kappa$ (cf.
Sec.~\ref{sec:PD_high_kappa}).
         Inhibitor-mediated failure is present in (a), direct failure in (b). 
         Marked with a thin solid red line in the top panel of (a) and (b) is the steady state value of the inhibitor.
         Other parameters as in Fig.~\ref{fig:unperturbed_timeseries}.}
\label{fig:PD_high_kappa}       
\end{figure*}

If the coupling strength $\kappa$ gets close to the maximum value $\kappa_{\text{max}}\approx 4.75$ for $N=150$ and
$k=2$, for the majority of values $d$ we find propagation failure.
In this regime we identify two mechanisms of propagation failure similar to the case of very low $\kappa$ but for
different reasons (Fig.~\ref{fig:PD_high_kappa}).

Firstly, shortcut blocking is active in this region of the coupling strength $\kappa$, therefore no secondary wave pairs
are excited at all.
If the coupling strength is very high, the excited part of the wave occupies many nodes.
For the inhibitor there is not enough time or space to relax completely to the rest value as in the case of low coupling
strengths.
Thus, the inhibitor level is above the rest value everywhere on the ring, making the wave more vulnerable.

For inhibitor-mediated failure (Fig.~\ref{fig:PD_high_kappa}(a)), a subthreshold excitation triggered through the
additional link leaves the inhibitor level at the other end of the link even more elevated.
This leads to suppression of the excitation at this node when the wave arrives and thus to propagation failure.
As in the low-$\kappa$ case, this mechanism is dependent on the distance $d$ of the additional link.

When $\kappa$ comes even closer to the maximum value, the backcoupling to an unexcited node is enough to suppress
propagation (Fig.~\ref{fig:PD_high_kappa}(b)).
The wave will fail to propagate at the first end of the additional link that is reached by the wave.
Here, no dependence on $d$ is present except that it needs to be larger than the width of the excited part of the wave.

We also observe solutions that show quenched propagation by an intermediate form of these two mechanisms. 
Moreover, in inhibitor-mediated failure in this regime of $\kappa$ there are also solutions with suppressed propagation
in which the inhibitor level at the second end of the link builds up over several periods of the wave until the
propagation is quenched.

$\kappa_\text{max}$ over almost all values of $d$, direct failure is the principal mechanism.
long-range link distances $d$ around 30.

\subsection{Phase Diagrams}\label{sec:phase_diags}
As already mentioned in Sec.~\ref{sec:spat_temp_patt}, the solutions on the minimal small-world network either decay to
the stable homogeneous state or show ongoing spatio-temporal activity with a well defined period $T_d(\kappa)$.
In Sec.~\ref{sec:spat_temp_patt} we have defined the relative period as $\vt_d(\kappa) \equiv T_d(\kappa)/T_0(\kappa)$,
where $T_0(\kappa)$ is the period that a traveling pulse on the regular ring shows for coupling strength $\kappa$.
From the relative period we can directly infer the influence that the additional link with a distance $d$ has.
In order to capture all observed spatio-temporal behavior in one quantity, we assign the value $-1$ to the relative
period, when wave propagation is quenched, or  mark the plot of $\vt_d(\kappa)$ with a special color at these points.

\begin{figure*}
\resizebox{1.0\textwidth}{!}{%
  \includegraphics{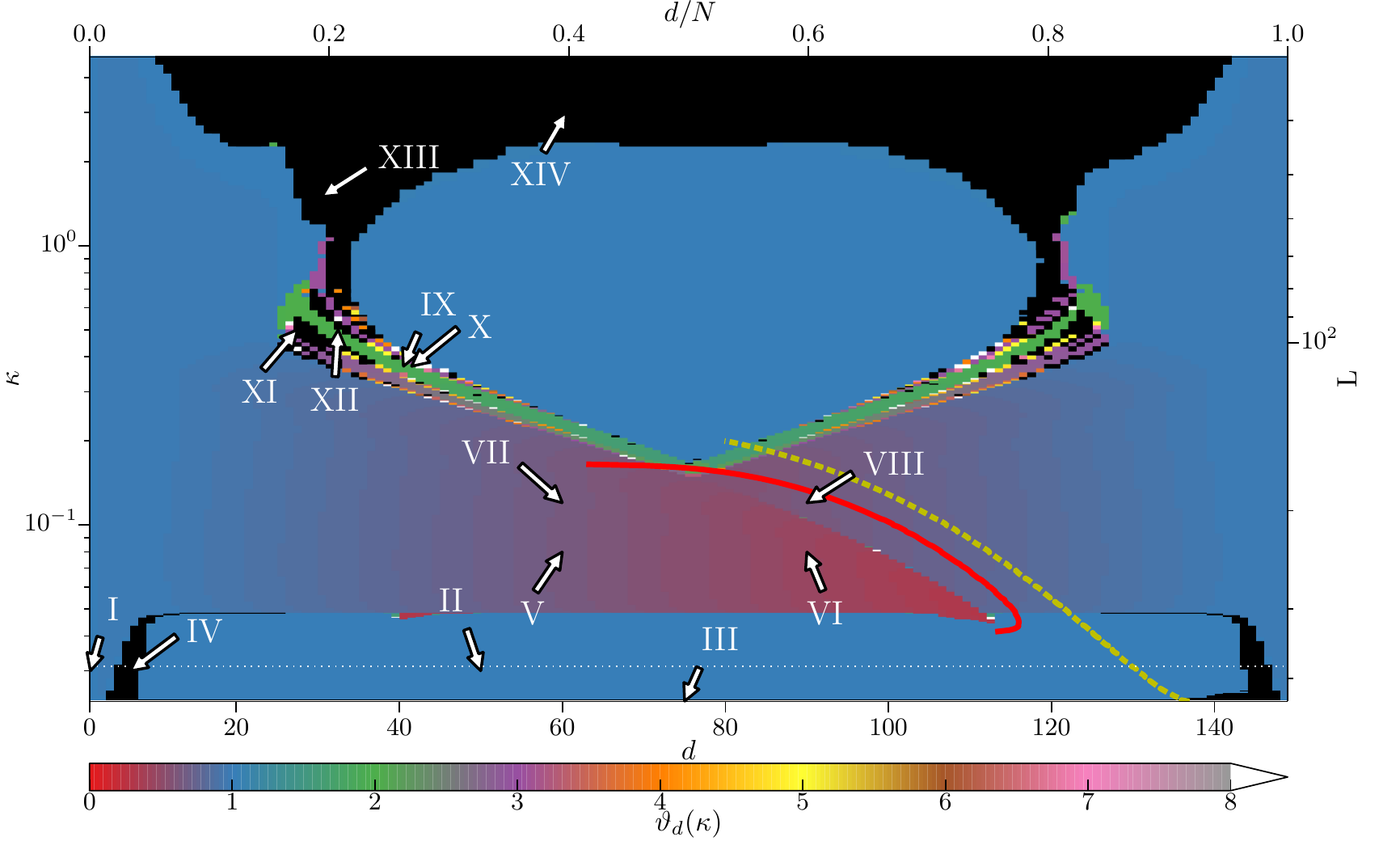}
}%
\caption{Color-coded relative period $\vartheta_d(\kappa)$ in the $(d,\kappa)$ plane. 
         Right vertical and upper horizontal axes show $L$ and $d/N$. 
         Black regimes indicate propagation failure.
         Yellow dashed line shows analytic approximation of pulse width without corrections, 
         red solid line shows analytic approximation to pulse width including corrections (cf.
Sec.~\ref{sec:analytics}).
         The white dotted line gives the analytically estimated value for $\kappa_2$.
         For the points marked by an arrow, timeseries are shown:
         I,II:     Fig.~\ref{fig:unperturbed_timeseries}(a,b),
         III,IV:   Fig.~\ref{fig:direct_and_indirect_failure}(a,b),
         V,VI:     Fig.~\ref{fig:period_decrease_and_bistability}(a,b),
         VII,IIX:  Fig.~\ref{fig:shortcut_block}(a,b),
         IX,X:     Fig.~\ref{fig:period_multiplication}(a,b),
         XI,XII:   Fig.~\ref{fig:PD_complex}(a,b),
         XIII,XIV: Fig.~\ref{fig:PD_high_kappa}(a,b).
         Other parameters as in Fig.~\ref{fig:unperturbed_timeseries}.}
         \label{fig:phase_diag_150_2_full}
\end{figure*}

A density plot of $\vt_d(\kappa)$ for fixed $N$ and $\kappa$ summarizes the occurrence of the spatio-temporal phenomena
described in Sec.~\ref{sec:spat_temp_patt} very clearly.
For $N=150$ and $k=2$, this `phase diagram' is shown in Fig.~\ref{fig:phase_diag_150_2_full}.
The values of the examples discussed in Sec.~\ref{sec:spat_temp_patt} are marked by Roman numerals.
Note that this diagram shows an almost perfect mirror symmetry with respect to $d$.
An asymmetry occurs only in the region of period decrease and shortcut blocking, that is, the region that becomes
increasingly red with increasing $d$ between $\kappa\approx 0.05$ and $\kappa\approx 0.2$.
From now on we use the term `phase diagram' for a density plot of $\vt_d(\kappa)$ at fixed values of $N$ and $k$.
Note that the diagram Fig.~\ref{fig:phase_diag_150_2_full} extends over $d\in\{k+1,...,N-(k+1)\}$.
Thus, by the shift symmetry discussed in Sec.~\ref{sec:model}, the asymmetry between $d$ and $\tilde d=N-d$ is due to
different initial conditions for the same $d$.

\subsubsection{Regimes of the coupling strength}\label{sec:limiting_regimes}
\paragraph{(i) Small $\kappa$ -- discrete regime}
When the coupling strength $\kappa$ is very small, the propagation takes place in a saltatory way, i.e., only one nodes
is fully excited at one instance of time.
In this regime of $\kappa$ -- provided that $N$ is large enough -- we observe no dispersion effects, where the wave
reaches nodes again before they have relaxed into the steady state.

The main effects in this regime are direct failure and inhibitor-mediated failure (cf. Secs.~\ref{sec:PD_direct} and
\ref{sec:PD_inh_med}, Fig.~\ref{fig:direct_and_indirect_failure} and points III and IV in
Fig.~\ref{fig:phase_diag_150_2_full}).
These effects are purely local in the sense that the size $N$ of the network has no influence on the values of $d$ and
$\kappa$ where these phenomena occur.
The values $\kappa_{\text{min}}$ below which no wave propagation is possible even on the regular ring, $\kappa_1$ below
which any distance $d>0$ of the additional link leads to (direct) failure, and $\kappa_2$ below which certain link
distances $d$ lead to (inhibitor mediated) failure and above which the generation of secondary wave pairs becomes
possible are shown in Table~\ref{tab:kappa_vals}.

\paragraph{(ii) High $\kappa$ -- continuum regime}
At high values of the coupling strength $\kappa$, the propagation takes place in a continuous fashion, i.e., at one
instance of time many nodes are in the excited state and more than one node is in the transition regime between excited
and non-excited state.
Moreover, for the highest values of $\kappa$, dispersion effects play a role, as, e.g., described in
Sec.~\ref{sec:PD_high_kappa}.

For this reason it is advantageous to take a different viewpoint.
In the limit $N\to\infty,\ \kappa\to\infty$ with 
\begin{align}
  \frac{N}{\sqrt{\kappa q(k)}} =\text{const} \equiv L\,, \label{eq:definition_L}
\end{align}
the system on the regular ring, Eq.~\eqref{eq:dynamics} becomes a continuous reaction-diffusion system with periodic
boundary conditions and domain size $L$, $q(k)=1/6 k(k+1)(2k+1)$ as discussed in Sec.~\ref{sec:spat_temp_patt}.
Considering a finite regular ring as the discretization of such a reaction-diffusion system, we can use $d/N$ instead of
$d$, and $L$ instead of the coupling strength $\kappa$ as quantities to compare networks with different $N$ and $k$ at
high coupling strengths.
coordinates $L$ and $d/N$.

\begin{figure*}
\resizebox{1.0\textwidth}{!}{%
  \includegraphics{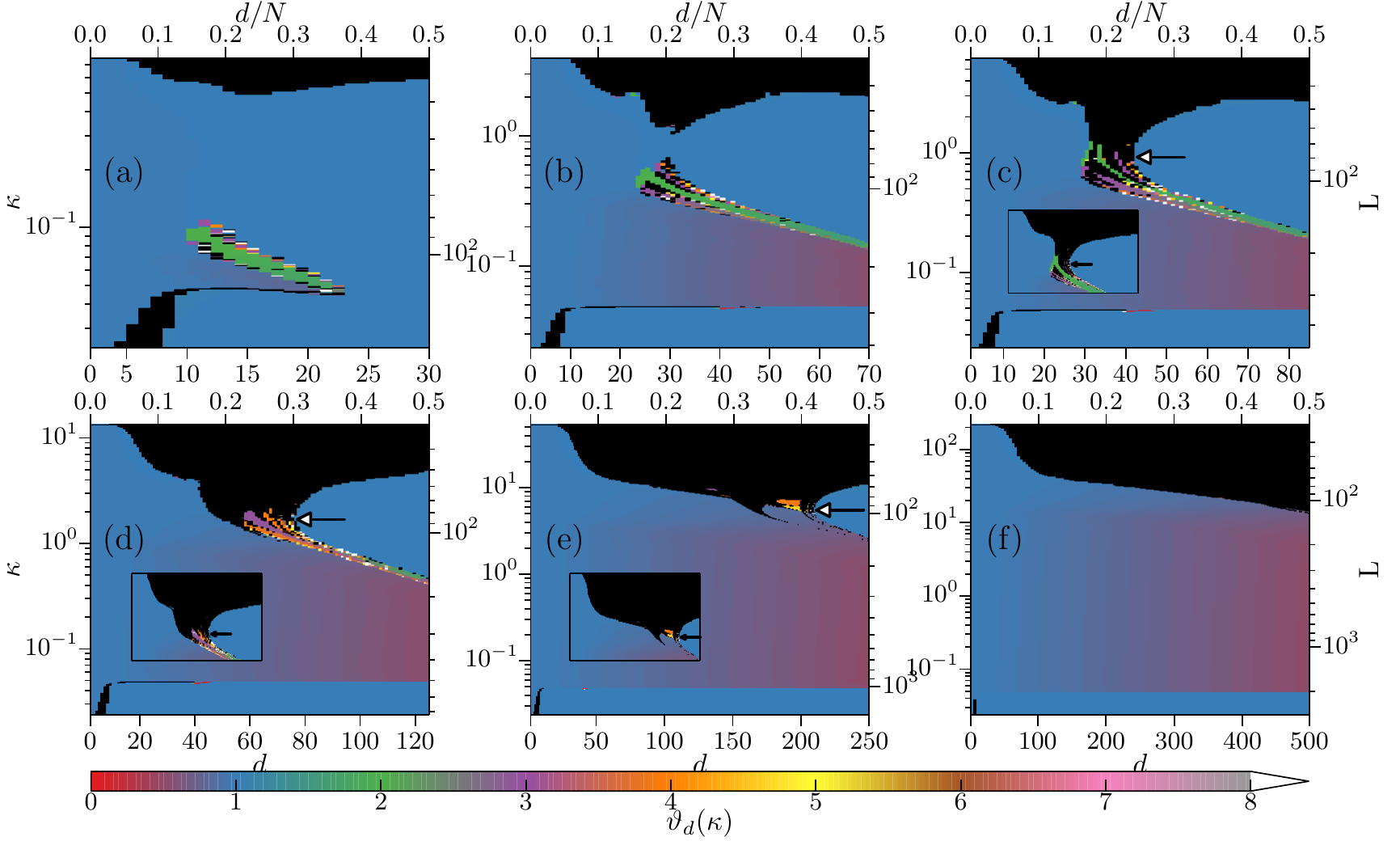}
}%
\caption{Typical evolution of phase diagrams with increasing $N$. 
         Color-coded $\vartheta_d(\kappa)$ in the $(d,\kappa)$ plane. 
         (a) $N\!=\!60  ,\,k\!=\!2$,
         (b) $N\!=\!140 ,\,k\!=\!2$,
         (c) $N\!=\!170 ,\,k\!=\!2$ (inset $N\!=\!510,\, k\!=\!3$),
         (d) $N\!=\!250 ,\,k\!=\!2$ (inset $N\!=\!700,\, k\!=\!3$),
         (e) $N\!=\!500 ,\,k\!=\!2$ (inset $N\!=\!1400,\,k\!=\!3$),
         (f) $N\!=\!1000,\,k\!=\!2$.
         Right and upper axes show $L$ and $d/N$ as units of measurement.
         Black regions indicate propagation failure.
         Other parameters as in Fig.~\ref{fig:unperturbed_timeseries}.}
       \label{fig:phase_diagrams_increase_N}
\end{figure*}

Both types of invariance (small $\kappa$ and high $\kappa$) can be seen in Fig.~\ref{fig:phase_diagrams_increase_N},
where for $k=2$, the phase diagram is shown for six different, increasing values of $N$.
The insets with phase diagrams for $k=3$ are discussed in Sec.~\ref{sec:scaling_behavior}.
The $\kappa$ and $d$ values of the horn-like structure in the lower left of each panel stay the same to a very good
approximation.
This is the regime of inhibitor-mediated failure as described in Sec.~\ref{sec:PD_inh_med}.
The $L$ and $d/N$ values (shown on the right vertical and upper horizontal axes) for the left border of the direct
failure regime also remain (approximately) constant as well as the maximum value of $L$ that is attained at all.
Also here, at the lowest values of $L$, the network size $N$ has only very little influence.

\paragraph{(iii) Intermediate $\kappa$}
In the region between the two limiting regimes, the appearance of the phase diagram changes drastically when $N$ (or
$k$) is changed.
Most of the complex behavior explained in Sec.~\ref{sec:spat_temp_patt}, like, e.g., period multiplying takes place in
this intermediate regime of the coupling strength.

The mechanisms behind these phenomena can be regarded as a well balanced interplay between 
\begin{inparaenum}[(i)]
\item    the discrete network nature of the system,
\item    the transition to the continuum regime and
\item    beginning dispersion effects.
\end{inparaenum}
The discrete network nature of the system  allows for triggering of secondary wave pairs through the additional
long-range link.
Due to the transition to the continuum regime, where many nodes are excited at the same time,
\begin{inparaenum}[a)]
   \item [(a)]     the node at one end of the additional link is perturbed longer and thus more strongly when an
excitation passes the other end, and 
   \item [(b)]     excitation is not lost easily when one single node is perturbed through the additional link.
\end{inparaenum}
Beginning dispersion effects cause the inhibitor level to be still elevated at the other end of the shortcut from the
previous wave passage.

The network size $N$ has an impact on two out of this list of three items and thus must clearly influence the appearance
of the phase diagram.
This is demonstrated in Fig.~\ref{fig:phase_diagrams_increase_N}.
In Fig.~\ref{fig:phase_diagrams_increase_N}(f), $N=1000$ and period multiplying is not observed anymore.
For even larger $N$ the appeareance of the phase diagram does not change significantly any more.
Thus for fixed nearest neighbor number $k$ we have a family of phase diagrams parameterized by the size of the network
$N$.

\begin{figure*}
\resizebox{1.0\textwidth}{!}{%
  \includegraphics{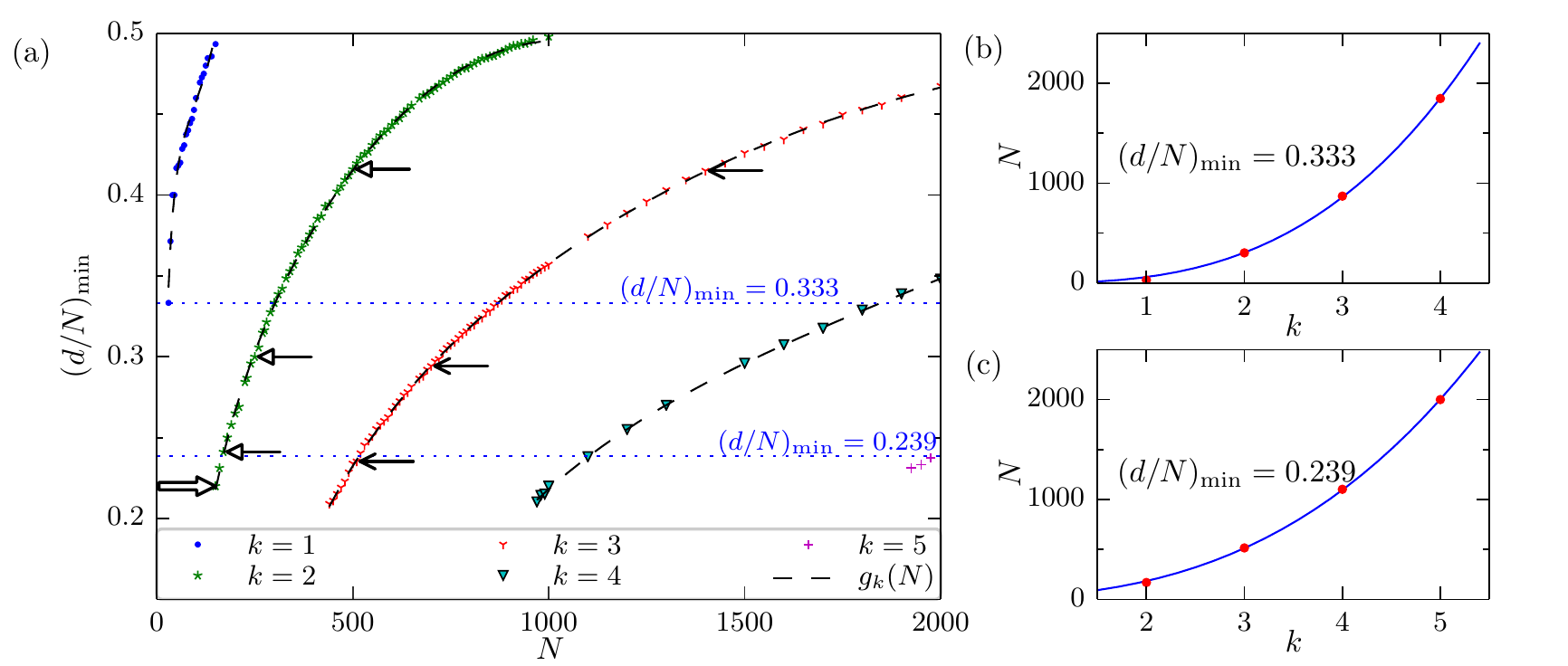}
}%
\caption{Scaling behavior of intermediate region of phase-diagrams.
         (a) Value $(d/N)_\text{min}$ at characteristic point introduced in Sec.~\ref{sec:scaling_behavior} and marked in Fig.~\ref{fig:phase_diagrams_increase_N}(c-e) vs. $N$ for $k=1,2,3,4,5$. Dashed line for fitted curves.
         (b)  $N$ values for $(d/N)_{\text{min}}=0.333$ vs. $k$, measured data points (red dots) and fit to const$\cdot q(k)$ (blue line).
         (c)  Same as (b) for $(d/N)_{\text{min}}=0.239$.
         Arrows in (a) mark data points for Fig.~\ref{fig:phase_diagrams_increase_N}(c)-(e) (white tip), insets of Fig.~\ref{fig:phase_diagrams_increase_N}(c)-(e) (black tip) and Fig.~\ref{fig:phase_diag_150_2_full} (white tip and stem). Other parameters as in Fig.~\ref{fig:unperturbed_timeseries}.
       }
       \label{fig:scaling_behavior}
\end{figure*}

\subsubsection{Scaling behavior of the intermediate regime}\label{sec:scaling_behavior}
The nearest neighbor number $k$ changes the balance between the influence of the local links (those that are present in the regular ring) and the additional link of the minimal small-world model. 
Increasing $k$ effectively lowers the weight of the additional link, by attaching more local links to every node.
Also, in the definition of $L$, Eq.~\eqref{eq:definition_L}, increasing $k$ has the same effect as increasing $\kappa$.
Thus it influences all three items (i)-(iii) of the above list, and thus has an impact upon the phase diagrams as well as $N$.

Next, we discuss how the influence of $k$ and $N$ on the phase diagrams relate to each other.
Given a phase diagram for $(N,k)$, it is possible to find higher values $(\tilde N,\tilde k)$ at which the phase diagram looks very similar.
In order to quantify this, we choose a characteristic point which is present in the `middle phase' of the evolution of phase diagrams.
We consider the border between propagation failure to lower values of $d$ and sustained wave propagation without influence of the additional link to higher values of $d$ at intermediate to high coupling strengths.
At some value of $\kappa$, this border attains a minimum value of $d_{\text{min}}$ or $(d/N)_{\text{min}}$, respectively.
This point is marked by an arrow in Figs.~\ref{fig:phase_diagrams_increase_N}(c)-(e).

Choosing a fixed $k=1,2,3,4,5$, we confirm the existence and determine the position $(d/N)_{\text{min}}$ of this point with increasing $N$.
The obtained values are shown in Fig.~\ref{fig:scaling_behavior}.
The measured data $(d/N)_{\text{min}}$ in dependence on $N$ can be fit to a function $g_k(N) = a_kN^2+b_kN+c_k+d_kN^{-1}$ to very good approximation.
So far we do not have a good explanation or motivation for this particular form of the fitting function.

In a second step, we choose a fixed value for $(d/N)_{\text{min}}$ and use the inverse of the fitting functions $g_k(N)$ to obtain the value of $N$ at which this desired value of $(d/N)_{\text{min}}$ is attained for a certain $k$.
This leaves us with $(N,k)$ tuples $(N_1,1),...,(N_5,5)$ for which the phase diagrams show the same value of $(d/N)_{\text{min}}$.
Moreover, these phase diagrams with the same value of $(d/N)_{\text{min}}$ also show other similar features at similar locations (measured in $L$ and $d/N$), though the diagrams are not completely identical.
The insets in Figs.~\ref{fig:phase_diagrams_increase_N}(c)-(e) show the phase diagrams for networks with the same $(d/N)_{\text{min}}$ but different $N$ and $k=3$ for illustration.
The corresponding data points in Fig.~\ref{fig:scaling_behavior} are marked by an arrow, as is the data point for Fig.~\ref{fig:phase_diag_150_2_full}.

At the values $(d/N)_{\text{min}}$ which we have examined, we find that the points $(N_1,1),...,(N_5,5)$ lie on a curve described by $N/q(k) = \text{const}$.
This is shown exemplarily for $(d/N)_{\text{min}}=0.333$ and $(d/N)_{\text{min}}=0.239$ in Figs.~\ref{fig:scaling_behavior}(b),(c).

This behavior is interesting because in the limit $N\to\infty,\ \kappa\to\infty$ with $L$ from Eq.~\eqref{eq:definition_L}, the domain length $L$ of the limiting continuous system $\partial_{t} u = u-\frac{u^3}{3}-v + \partial_{xx}u,\ \partial_{t} v=\ve(u-\beta), u(t,0)=u(t,L)$, $v(t,0)=v(t,L)$ and with it all other length scales depend on $N/\sqrt{q(k)}$ instead of $N/q(k)$.
As explained above, the behavior in the intermediate regime can be viewed as an interplay between discrete and continuum effects, thus it is surprising that in the complicated intermediate regime such a simple scaling law can be found at all.

\subsection{Analytics}\label{sec:analytics}
\subsubsection{Approximation of critical distance for shortcut blocking}\label{sec:appr-crit-dist}
The triggering of a secondary wave pair through the additional link can be prevented by the inhibitor level at this node being too high from the last passage of the wave.
We make an approximation of the critical distance $d_{\text{crit}}$ up to which triggering of a secondary wave pair is prevented.


We consider Eqs.~\eqref{eq:dynamics} at the node at which the secondary wave pair would be triggered through the additional link.
In the limit of  $\ve\to 0$, the transition to the excited state of this node is described by the fast subsystem of Eqs.~\eqref{eq:dynamics}.
This is just Eq.~\eqref{eq:dynamics_v} with $v$ acting as a parameter.
This node receives input from its $2k$ neighbors on the regular ring and from the excited node at the other end of the additional link.
The neighboring nodes of this node as well as the neighboring nodes of the node at the other end of the additional link are approximately in the same state so that the coupling term in their equation vanishes and their $u$-values can be described by the (stable) fixed points of $\dot u = u - u^3/3 -v$.
We denote those fixed points by $z_1(v) < z_2(v) <z_3(v)$, where $z_1$ and $z_3$ are stable.
Thus the approximating bistable system for the node under consideration is
\begin{align}
  \dot u &= u - \frac{u^3}{3} - v + \kappa \left(  z_3(v_0) + 2kz_1(v) - (2k+1)u \right),     \label{eq:bist_sys_at_trigger_node}
\end{align}
where $v_0=\beta-\beta^3/3$ is the (homogeneous) steady state value of the inhibitor in Eq.~\eqref{eq:dynamics}.

For $v_0 \le v < v_{\text{crit}}$, Eq.~\eqref{eq:bist_sys_at_trigger_node} possesses only one stable fixed point which is the excited value. 
Another stable and one unstable fixed point appear in a saddle-node bifurcation at $v=v_{\text{crit}}$ so that for $v>v_{\text{crit}}$ the node can remain at rest.
The location of the saddle-node bifurcation is determined by the solution $(u^*,v_{\text{crit}})$ of 
\begin{align}
  0 &= u - \frac{u^3}{3} - v + \kappa \left(  z_3(v_0) + 2kz_1(v) - (2k+1)u \right)   \nonumber \\
  0 &= 1 - \kappa (2k+1) - u^2.                                                        \label{eq:sn_point}
\end{align}
From the second equation we infer that this approximation can only be used up to $\kappa=1/(2k+1)$.
The reason why this approximation breaks down for larger $\kappa$ is that the coupling term in the unexcited neighboring nodes becomes stronger with increasing $\kappa$ and thus their state cannot be approximated by $z_1(v)$ any more.

In a second step, we approximate the time $t_{\text{crit}}$ it takes until a node reaches this critical inhibitor level after becoming excited.
In this approximation, we neglect the time that the transitions to and from the excited state take, only taking into account the time it spends on the slow manifold ($u$-nullcline) of Eqs.~\eqref{eq:dynamics}.
While on the slow mainfold, a node has approximately the same $u$ value as its neighbors and we can again neglect the coupling term.
Performing the transformation $t\to \tilde t = t/\ve$, and taking the limit $\ve\to 0$, from Eqs.~\eqref{eq:dynamics} we get
\begin{align}
  \dot u &= 0 = u-\frac{u^3}{3} - v     \nonumber \\
  \dot v &= u-\beta .                     \nonumber 
\intertext{Inserting the algebraic constraint of the first equation into the second gives}
  \dot u &= \frac{u-\beta}{1-u^2}   \label{eq:slow_eq},
\end{align}
where all derivatives are with respect to $\tilde t$.
From this equation we can get $t_{\text{crit}}$ by integrating
\begin{align}
  t_{\text{crit}} &= \ve^{-1}\left (\int_{z_3(v_0)}^1 + \int_{-2}^{z_1(v_{\text{crit}})} \right) \frac{1-u^2}{u-\beta} \text{d} u \label{eq:delta_t},
\end{align}
where the boundaries of the integral are the $u$-values on the $u$-nullcline that the system jumps to and from in the limit $\ve\to0$.

With that, $d_{\text{crit}}/N = t_{\text{crit}} / T_0$, where we obtain $T_0$ from the simulations for $d=0$.
The calculated value for $1-d_{\text{crit}}/N$  is shown as a yellow dashed line in Fig.~\ref{fig:phase_diag_150_2_full}.

The main reason for the deviation is that close to the saddle-node bifurcation of Eq.~\eqref{eq:bist_sys_at_trigger_node}, it takes a finite time until $u$ has reached full excitation.
The node that triggers the excitation through the additional link however is only excited for a finite time which need not be sufficient to lead to a full excitation of $u$. 

We take this into account by calculating the saddle-node normal form $\dot y = a (v-v_{\text{crit}}) + b y^2$, with $y\equiv u-u^*$ for  Eq.~\eqref{eq:bist_sys_at_trigger_node} with
\begin{align}
  a &= 1 + \frac{2k\kappa}{1-z_1(v_{\text{crit}})^2}\ ,  &  b &= -u^* \nonumber
\end{align}
Close to the bifurcation point, the time $t_{\text{exc}}$ until excitation can be estimated by the time, the variable $y$ of the normal form needs to go from $-\infty$ to $+\infty$.
\begin{align}
  t_{\text{exc}} &= \frac {\pi}{\sqrt{ab(v-v_{\text{crit}})}}\,.
\end{align}
Setting this equal to the time the excitation lasts that is, first integral in Eq.~\eqref{eq:delta_t}, we obtain 
\begin{align*}
  v &= v_{\text{crit}} + \frac{1}{ab}\left( \frac{\ve\pi}{\int_{z_3(v_0)}^1 \frac{1-u^2}{u-\beta} \text{d} u} \right)^2 \,.
\end{align*}
In Fig.~\ref{fig:phase_diag_150_2_full}, $1-d_{\text{crit}}/N$ for the solution of this equation is plotted as a red solid line.

\subsubsection{Analytical approximation for $\kappa_2$}\label{sec:analyt-appr-kappa2}
We can use a very similar technique to estimate $\kappa_2$.
This coupling strength marks the value below which no secondary wave pairs can be triggered through the additional link.
Instead of treating $v$ as the bifurcation parameter in Eq.~\eqref{eq:bist_sys_at_trigger_node}, we use $\kappa$, fixing $v=v_0$.
However, the value $\kappa^*$ for the saddle-node bifurcation, attained by this calculation is much too small.
As before, the main reason is the finite time, the excited node can exert its influence through the additional link.
Following the steps of Sec.~\ref{sec:appr-crit-dist}, we obtain the estimate
\begin{align*}
  \kappa_2^{\text{est}} = \kappa^* + \frac{1}{ab}\left( \frac{\ve\pi}{\int_{z_3(v_0)}^1 \frac{1-u^2}{u-\beta} \text{d} u} \right)^2 \,,
\end{align*}
where we now have
\begin{align*}
  a &= 2k\beta + z_3(v_0) - (2k+1) u^* \,,  &  b &= -u^*\,.
\end{align*}
For $k=2$, $\kappa_2^{\text{est}}\approx 0.031$, which is still too small (cf. Table~\ref{tab:kappa_vals}).
This value is marked by a dotted white line in Fig.~\ref{fig:phase_diag_150_2_full}.

The deviation of the approximate values for $\kappa_2$ and $d_{\text{crit}}$ are in the same direction. Possible reasons include the following:
\begin{inparaenum}[(i)]
  \item $\ve$ is not exactly zero, and thus the excitation does not happen instantaneously. 
Therefore, the inhibitor level at the node that is about to become excited will rise to some extent.
A merely partially excited node at the other end of the additional link leads to a sub-threshold excitation. 
This in turn makes a larger global coupling strength necessary to trigger a full excitation.
  \item The backcoupling through the additional link makes the excitation of the node that triggers the secondary wave pair smaller.
\end{inparaenum}

\section{Conclusion}\label{sec:conclusion}
In this paper we have presented a detailed investigation of spatio-temporal patterns on a minimal case of a Newman-Watts small-world network model. 
This minimal case consists of regular ring with one additional long-range link.
When the distance $d$ that this link bridges is taken into account as a parameter, the topology of this system comprises no randomness in contrast to general Newman-Watts small-world networks that have been studied in Ref.~\cite{ISE14}.

We have shown that starting from a simple traveling wave solution, introducing the additional link induces many different intriguing spatio-temporal patterns with largely varying temporal periods.
The specific pattern depends on the global coupling strength $\kappa$ and the distance $d$.
We have systematically discussed the mechanisms behind the different patterns.
By comparing the period of each spatio-temporal pattern with the period of a traveling wave on a regular ring, we have provided a comprehensive overview of the behavior of a network of given size $N$ and nearest neighbor number $k$  for varying $d$ and $\kappa$ in a single diagram.

We have shown that such a phase diagram changes significantly when either $N$ or $k$ is varied. 
However, by changing $N$ we have identified a general trend of these diagrams, which is independent of $k$.
For very low coupling strengths, only $k$ plays a role for the appearance of these diagrams whereas in the regime of high coupling strength, transformed coordinates related to a continuum description provide an invariant description.
By systematically studying points on phase diagrams for different $N$ and $k$, we have found that diagrams with the same value of  $N/(k(k+1)(2k+1)$ resemble each other very closely for intermediate coupling strengths.
To the best of our knowledge, this scaling behavior has not been reported before. 

Our results contribute to a better understanding of pattern forming dynamic networks.
Due to the simplicity of our system, the presented results are expected to be very useful in the design of networks that can generate diverse patterns with different periods.
The control of pattern selection in our system can be performed with very little effort by just modifying either the global coupling strength or the distance of a single link.
Therefore, controlling complex spatio-temporal behavior does not require complex topologies.
A single, well-placed shortcut is sufficient.

\section*{Acknowledgements}
This work was supported by DFG in the framework of SFB 910. PH acknowledges support by BMBF (grant no. 01Q1001B) in the framework of BCCN Berlin (Project A13). We thank Judith Lehnert for helpful discussions.


\end{document}